# A Dual-Feature Extractor Framework for Accurate Back Depth and Spine Morphology Estimation from Monocular RGB Images

Yuxin Wei, Yue Zhang, Moxin Zhao, Chang Shi, Jason P.Y. Cheung, Teng Zhang, *IEEE Senior Member,* Nan Meng[†], *IEEE Member*

*Abstract*—Scoliosis is a prevalent condition that impacts both physical health and appearance, with adolescent idiopathic scoliosis (AIS) being the most common form. Currently, the main AIS assessment tool, X-rays, poses significant limitations, including radiation exposure and limited accessibility in poor and remote areas. To address this problem, the current solutions are using RGB images to analyze spine morphology. However, RGB images are highly susceptible to environmental factors, such as lighting conditions, compromising model stability and generalizability. Therefore, in this study, we propose a novel pipeline to accurately estimate the depth information of the unclothed back, compensating for the limitations of 2D information and then estimate spine morphology by integrating both depth and surface information. To capture the subtle depth variations of the back surface with precision, we design an adaptive multiscale feature learning network, named Grid-Aware Multiscale Adaptive Network (GAMA-Net). This model uses dual encoders to extract both patch-level and global features, which are then interacted by the Patch-Based Hybrid Attention (PBHA) module. The Adaptive Multiscale Feature Fusion (AMFF) module is used to dynamically fuse information in the decoder. As a result, our depth estimation model achieves remarkable accuracy across three different evaluation metrics, with scores of nearly 78.2%, 93.6%, and 97.5%, respectively. To further validate the effectiveness of the predicted depth, we integrate both surface and depth information for spine morphology estimation. This integrated approach enhances the accuracy of spine curve generation, achieving an impressive performance of up to 97%.

*Clinical Relevance*—Our approach integrates a novel depth estimation model with RGB images to provide comprehensive and accurate spine morphology analysis.

## I. Introduction

Scoliosis is a common three-dimensional (3D) spinal deformity characterized by lateral curvature in the coronal plane, often accompanied by vertebral rotation or segmental bending [1]. It can be classified into neuromuscular, congenital, and idiopathic types [2]. Adolescent Idiopathic Scoliosis (AIS) is the most prevalent [3], which usually develops during children's and teenagers' fast development phases, especially between the ages of 10 and 18 [4]. Even though AIS is often asymptomatic in its early stages, it can have long-term effects on patients' health as they age [5]. In particular, when the spinal curvature exceeds an essential threshold, AIS can worsen over time and cause observable physical abnormalities [6]. A number of serious health problems, such as cardiovascular dysfunction, lung impairment, and restricted physical activity, can arise from untreated AIS, greatly impairing health and quality of life [4, 7].

Full-spine radiographic examination remains the clinical gold standard for diagnosing and managing scoliosis [8]. A typical scoliosis patient undergoes approximately 22 radiographic examinations over a three-year treatment period [9]. Children are particularly vulnerable to radiation exposure due to their higher rate of cells division, which significantly increases their risk of developing cancer [8, 9]. Moreover, AIS also has a higher prevalence in females, with incidence ratios ranging from 1.5:1 to 3:1, compared to males [4]. Female patients, in particular, face an elevated rise of radiation-induced breast cancer due to increased exposure [9]. In addition, radiographic equipment is typically large and expensive, making it difficult to perform in remote areas with limited medical resources. This poses a significant barrier to the widespread adoption of radiographic screening, limiting its viability as a scalable solution for scoliosis detection.

To address these challenges, the development of radiation-free scoliosis screening methods and the miniaturization of diagnostic equipment have gradually become key research focuses in spinal health. With the rapid advancement of deep learning and the growing adoption of optical imaging technologies across various disciplines [10], [11], [12], an increasing number of studies have begun exploring AIS assessment through optical imaging approaches. For example, Zhang et al. [13] proposed a method for assessing the severity of AIS by analyzing smartphone photographs of unclothed backs with a deep learning model. Zhu et al. [14] also explored scoliosis severity grading and Cobb angle estimation using natural RGB images. However, these studies relying on RGB images focus primarily on coarse spine analysis as RGB images are limited to 2D information and are highly susceptible to factors such as lighting conditions. Such

Research supported by Department Seed Fund of The University of Hong Kong (200011026), Seed Fund for Basic Research for New Staff of The University of Hong Kong (103034011), and The National Natural Science Foundation of China Young Scientists Fund (NSFC 82402398).

Yuxin Wei, Yue Zhang, Moxin Zhao, Chang Shi, Jason P.Y. Cheung, Teng Zhang, Nan Meng. Authors are with Department of Orthopaedics and Traumatology, The University of Hong Kong, Hong Kong.

e-mail: yxwei2002@connect.hku.hk, u3012167@connect.hku.hk, moxin@connect.hku.hk, chaseshi@hku.hk, cheungjp@hku.hk, tgzhang@hku.hk, nanmeng@hku.hk.

[†] Corresponding author

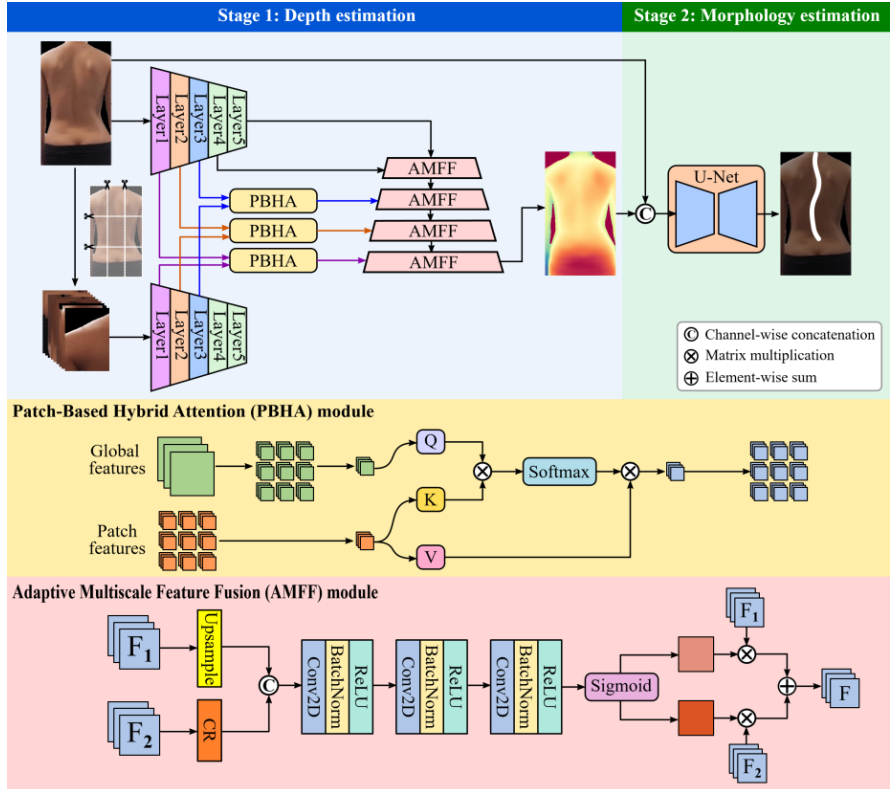

**Figure 1.** The overview of the proposed GAMA-Net model for depth estimation, including detailed illustrations of Patch-Based Hybrid Attention (PBHA) module and Adaptive Multiscale Feature Fusion (AMFF) module. The subsequent U-Net model is utilized for spine morphology estimation.

limitation prevents accurate and detailed assessments of spine morphology.

In recent years, researchers have incorporated depth information of the human back into their analyses, enabling more detailed assessments of spine morphology. For example, Kokabu et al. [15] utilized 3D depth data in convolutional neural networks (CNNs) for regression analysis, achieving precise predictions of the Cobb angle. Liang et al. [16] proposed a 3D spinal model reconstruction method using RGBD images of unclothed backs. Yang et al. [17] developed a method for measuring the Cobb angle and probe positioning by integrating ultrasound and RGBD data of the back. Unlike RGB images, depth provides valuable geometric information, which is essential for analyzing complex 3D structures. Relying solely on 2D planar data of human back, such as shape, size, color, and texture, is insufficient to comprehensively capture the intricate characteristics of the spine. However, specialized equipment, such as depth cameras, are expensive, limiting their accessibility in many settings. To address this limitation, we proposed a novel pipeline, designed to precisely predict depth information from standard RGB images. By integrating the surface features encoded in RGB images with the estimated geometric depth information, this approach can enhance the precision of spine morphology estimation.

Currently, depth estimation has been extensively studied due to its potential to improve spatial understanding [18, 19]. Compared to traditional scene depth estimation, the human back has a relatively narrow depth range (approximately 10 cm), exhibiting subtle depth variations. This imposes stricter requirements on the model, demanding the capability to capture these fine-grained depth fluctuations accurately. Although a few studies have incorporated depth into medical imaging tasks, for example, Chong et al. [20] proposed a dual-branch Siamese network for monocular RGB endoscopic images, Jeong et al. [21] introduced a three-step training method combining simulated data generation with simulation-to-reality transfer, Huang et al. [22] integrated depth estimation and segmentation tasks by proposing a unified model for laparoscopic images, and Khan et al. [23, 24, 25] proposed various face depth estimation methods, depth estimation specifically targeting human back images remains largely unexplored.

In this study, we proposed an innovative framework Grid-Aware Multiscale Adaptive Network (GAMA-Net). The framework accurately estimated the depth information of the unclothed human back from a single RGB image and then integrated the surface and depth information of the back to estimate the coronal spinal curve. To capture subtle variations in depth, we designed dual feature encoding framework, consisting of a patch-based feature extractor and global feature extractor, which were interconnected through the proposed Patch-Based Hybrid Attention (PBHA) module. This module effectively ensured the predicted depth maps capture fine-grained local variations while integrating global contextual information. Meanwhile, to further optimize feature integration in the decoder, we introduced an Adaptive Multiscale Feature Fusion (AMFF) module, which dynamically assigns weights to fused features. This approach effectively avoided information redundancy, resulting in a

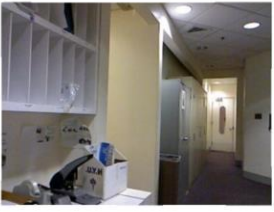 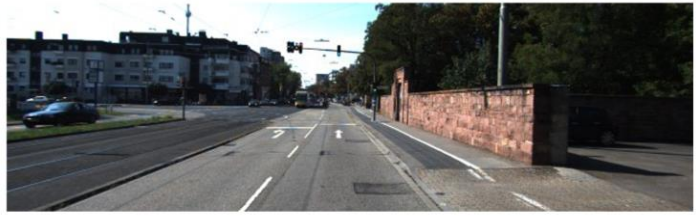 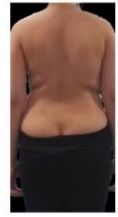

NYU dataset            KITTI dataset            Our dataset more adaptive and efficient feature fusion. To validate the utility of the generated depth information, we combined depth with the original RGB images to reconstruct spinal curves. These reconstructed curves offer clinicians' valuable insights into spine morphology, enhancing diagnostic accuracy.

**Figure 2.** The comparison of NYU, KITTI, and our unclothed back dataset.

The contributions of our study include:

1. We proposed a novel framework for depth estimation GAMA-Net, with dual feature extractors (patch-based feature extractor and global feature extractor), specifically designed to handle small variations in depth maps with high precision.

2. A hybrid attention module (PBHA module) and a feature fusion module (AMFF module) were proposed to improve feature extraction and fusion, respectively. The PBHA module strengthened the interconnection between the global and local patch-based features, ensuring a comprehensive representation of depth information. Meanwhile, the AMFF module can effectively combine high-level semantic information with low-level details, reducing redundancy during feature fusion.

3. To the best of our knowledge, this study is among the first to explore precise depth estimation of the human back using single RGB images, providing enriched geometric information to enhance spine morphology estimation. The proposed pipeline not only offers an advanced method for precise human back depth estimation but also has the potential to significantly improve the accuracy of spine morphology estimation compared to methods relying solely on RGB images.

## II. METHOD

### A. Dataset

The dataset was collected at Queen Mary Hospital (QMH) and The Duchess of Kent Children's Hospital (DKCH) in Hong Kong. It was approved by the Institution's Ethical Review Board. For each participant, a pair of unclothed back RGB and depth images, and a whole-spine standing posteroanterior radiograph were acquired. The dataset consisted of 2,213 images in total, with 1,619 patients (73.16%) who were female and 594 patients (26.84%) who were male. All participants were adolescents, aged between 10 and 18 years, with an average age of 16.02 years. Before data collection, participants have written informed consent. Exclusion criteria included individuals with psychological or systemic neurological disorders, congenital deformities, a history of spinal surgery, trauma affecting posture or mobility, and tumor-related conditions. The dataset was divided into training set, validation set, and testing set in a ratio of 8:1:1. The depth captured by the Azure Kinect DK camera served as the ground truth (GT) for depth estimation. Full-spine radiographs which from C7 to L5 were used as the GT for spine curve generation. The depth images and RGB images were pixel-aligned using Azure Kinect SDK 1.4.1, while the radiographs were aligned with the RGB images based on anatomical landmarks marked by spine specialists on image modalities.

### B. Pipeline

The proposed overall pipeline is illustrated in Figure 1, containing two stages, namely the depth estimation stage and the spine morphology estimation (curve generation) stage, along with a detailed integration of these two components.

#### 1. Data Pre-processing and Experimental Settings

To enhance the training performance of the analysis of spine morphology, a series of data augmentation techniques were applied to the input images including random horizontal flipping, random vertical flipping, and adding random Gaussian noise, which aimed to increase the diversity of the training dataset and improve the model's generalization capability. However, data augmentation techniques were not applied to the depth estimation stage due to potential disruptions in patch reassembly, a critical step in maintaining the integrity of the depth estimation process. To standardize the input for both stages, all images were resized to a resolution of 480×240 pixels prior to being fed into the models. The proposed model was implemented using the PyTorch framework and was trained on an NVIDIA RTX 3090 GPU. The learning rate was set as $10^{-4}$ during the whole training process, while a batch size of 16 was used for depth estimation and a batch size of 8 was used for curve generation.

#### 2. Depth Estimation Stage

Most existing depth estimation algorithms [26] are primarily designed and optimized for the scenarios with large depth range, such as NYU and KITTI datasets, showing in Figure 2. However, the depth variations are often extremely subtle in human back images, typically within just a few centimeters. Such small-scale depth differences pose new challenges to existing algorithms designed for scene-based depth estimation.

To capture subtle depth variations in human back images, we proposed a dual feature encoding framework. The first encoder, global feature extractor, processed the entire RGB images of the unclothed back to capture overall image features. The patch-based feature extractor, took as input multiple image patches obtained by dividing the original RGB image into a grid, extracting local fine-grained features from different regions. Considering that fine-grained features are predominantly captured in the shallow layers of the network [27], we proposed PBHA module, which facilitated interaction between global and fine-grained patch features of first three layers through a cross-attention mechanism. Specifically, the entire image was processed through the global feature extractor to obtain features $F_I$. Meanwhile, the input image $I \in R^{h \times w \times c}$ was divided into 9 equally size patches, denoted as $P_i$, where $i \in (0,8)$ and $P_i \in R^{\frac{h}{9} \times \frac{w}{9} \times c}$. Patch features $F_{P_i}$ were extracted from each patch with shared patch-based feature extractor, allowing the model to focus on localized details.

$$F_{P_i} = encoder(P_i) \quad (1)$$

$$F_I = encoder(I) \quad (2)$$

The PBHA module was used to further integrate global and patch information, as depicted in Figure 1. In this module, the global feature was divided into 9 patches $F_{I_i}$, where $i \in (0,8)$. Each $F_{I_i}$ was corresponding to the 9 patches from the input image. Cross-attention was then applied between each patch feature $F_{P_i}$ and its respective global feature $F_{I_i}$ to facilitate effective interaction and integration.

The working principle of cross attention in the PBHA was converting global features $F_{I_i}$ into $Query$ $(Q)$, and converting the patch features $F_{P_i}$ into $Key$ $(K)$ and $Value$ $(V)$. $W_Q$, $W_K$ and $W_V$ were the weight matrixes used to map features to different spaces.

$$Q_{I_i} = W_Q F_{I_i} \quad (3)$$
$$K = W_K F_{P_i} \quad (4)$$
$$V = W_V F_{P_i} \quad (5)$$

Attention weights were obtained by calculating the dot product of $Q$ and $K$ and then computed by $softmax$. We used the attention weight to perform weighted averaging on $V$ to get the final feature $F_{C_i}$ of each patch. The cross feature $F_C$ for entire image was then obtained by sum up all $F_{C_i}$.

$$A = softmax\left(\frac{Q_{I_i} K^T}{\sqrt{d_k}}\right) \quad (6)$$

$$F_{C_i} = AV \quad (7)$$

$$F_C = \sum_{i=0}^{8} F_{C_i} \quad (8)$$

Currently, several studies have demonstrated the superiority of multiscale features in image analysis [28, 29]. Traditional feature fusion method such as integrating features directly from different scales often leads to information redundancy [30], undermining the efficiency of feature fusion. To address this issue, we implemented a multiscale fusion with decoder strategy [23], named as Adaptive Multiscale Feature Fusion (AMFF) module. This module dynamically adjusted the fusion weights based on the importance of features at different scales, effectively avoiding redundant information while remaining critical details.

In AMFF module, the first step involved aligning the feature dimensions of different input feature maps. Specifically, the high-resolution feature $F_{high}$ extracted from the layer $i$ of the dual encoders was aligned the low-resolution feature $F_{low}$ extracted from the layer $i+1$, along with channel reduction of $F_{high}$. Once aligned, the two features were concatenated. A sigmoid activation function $\delta(\cdot)$ was then applied to the concatenated features to generate attention maps, which dynamically highlighted the most relevant regions for the task. Finally, the refined feature was obtained by combining the aligned feature maps obtained from dual encoders, which regard attention maps $w$ as the combining weight. Formally, the operations can be expressed as,

$$F_{concat} = concat(F_{aligned}^{high}, F_{aligned}^{low}, dim = 1) \quad (9)$$

$$w = \delta(Conv(F_{concat})) \quad (10)$$

$$F_{out} = w^{(0)} \times F_{aligned}^{high} + w^{(1)} \times F_{aligned}^{low} \quad (11)$$

In terms of evaluation metrics for depth estimation, they are typically divided into two categories: accuracy metrics and error metrics. The following are the equations about accuracy metrics:

$$Mean\ \delta < 1.25 = \frac{1}{N}\sum_{i=1}^{N} \delta(\max\left(\frac{\widehat{D}_i}{D_i}, \frac{D_i}{\widehat{D}_i}\right) < 1.25), \quad (12)$$

$$Mean\ \delta < 1.25^2 = \frac{1}{N}\sum_{i=1}^{N} \delta(\max\left(\frac{\widehat{D}_i}{D_i}, \frac{D_i}{\widehat{D}_i}\right) < 1.25^2), \quad (13)$$

$$Mean\ \delta < 1.25^3 = \frac{1}{N}\sum_{i=1}^{N} \delta(\max\left(\frac{\widehat{D}_i}{D_i}, \frac{D_i}{\widehat{D}_i}\right) < 1.25^3), \quad (14)$$

where $N$ is the total number of pixels, $\widehat{D}_i$ is the predicted depth value for the $i$-th pixel, $D_i$ is the GT depth value for the $i$-th pixel and $\delta(\cdot)$ is the indication function, which equals 1 if the condition is satisfied, otherwise 0.

The following equations are error metrics including absolute relative error (Abs Rel), root mean square error (RMSE), and squared relative error (Sq Rel).

$$Abs\ Rel = \frac{1}{N}\sum_{i=1}^{N} \frac{|D_i - \widehat{D}_i|}{D_i}, \quad (15)$$

$$RMSE = \sqrt{\frac{1}{N}\sum_{i=1}^{N}(D_i - \widehat{D}_i)^2}, \quad (16)$$

$$Sq\ Rel = \frac{1}{N}\sum_{i=1}^{N} \frac{(D_i - \widehat{D}_i)^2}{D_i}, \quad (17)$$

3. *Spine Morphology Estimation (Curve Generation) Stage*

The obtained precise depth maps were then utilized for accurate spine curve generation in the spine morphology estimation stage. This stage primarily aimed to evaluate the contribution of the depth maps, which were obtained in the depth estimation stage. The RGB images of the human back, captured during data collection, were concatenated with the estimated depth maps along the channel dimension to form RGBD image data. These RGBD images were then used as input to generate the spine curve.

**Table 1.** The evaluation of the effectiveness of PBHA module and AMFF module in the proposed GAMA-Net.

| Method | Error↓ | | | Accuracy↑ | | |
|---|---|---|---|---|---|---|
| | *Abs Rel* | *RMSE* | *Sq Rel* | *β<1.25* | *β<1.25²* | *β<1.25³* |
| Baseline | 0.2220 | 0.0684 | 0.0177 | 0.7406 | 0.9266 | 0.9675 |
| Baseline+ PBHA module | 0.2123 | 0.0662 | 0.0166 | 0.7621 | 0.9297 | 0.9674 |
| Baseline+ AMFF module | 0.2131 | 0.0654 | 0.0167 | 0.7705 | 0.9285 | 0.9666 |
| **GAMA-Net (with all modules)** | **0.1887** | **0.0638** | **0.0140** | **0.7815** | **0.9356** | **0.9747** |

*Abs Rel: absolute relative error, RMSE: root mean square Error, Sq Rel: squared relative error.

**Table 2.** The ablation study for evaluating the effectiveness of different components in the loss function.

| Loss | Error↓ | | | Accuracy↑ | | |
|---|---|---|---|---|---|---|
| | *Abs Rel* | *RMSE* | *Sq Rel* | *β<1.25* | *β<1.25²* | *β<1.25³* |
| $L_{grad}$ | 0.2251 | 0.0675 | 0.0186 | 0.7407 | 0.9140 | 0.9610 |
| $L_{SSIM}$ | 0.1969 | 0.0687 | 0.0151 | 0.7510 | 0.9304 | 0.9718 |
| $L_{Berhu}$ | 0.1868 | 0.0671 | 0.0142 | 0.7592 | 0.9309 | 0.9714 |
| $L_{Total}$ | **0.1887** | **0.0638** | **0.0140** | **0.7815** | **0.9356** | **0.9747** |

*Abs Rel: absolute relative error, RMSE: root mean square error, Sq Rel: squared relative error.

*A. Depth Estimation Loss*

In the depth estimation stage, we used BerHu loss (Reverse Huber Loss) [31] as the main optimization objective, which is

$$x = |\widehat{D} - D|, \quad (19)$$

$$L_{Berhu}(x) = f(x) = \begin{cases} |x|, & if\ |x| \leq c \\ \dfrac{x^2 + c^2}{2c}, & if\ |x| > c \end{cases}, \quad (20)$$

where $x$ is the difference between the predicted value $\widehat{D}$ and the true value $D$. The hyperparameter $c$ is to control the point at which the loss function switches within the error range. Due to the variation of spine is small, the threshold $c$ was empirically set to be 0.08.

This loss design combines the advantages of L1 loss and L2 loss. When the error is small, BerHu loss is equivalent to L1 loss, which can handle small range errors more robustly. When the error range is large, it amplifies the weight of large errors, thereby prompting the model to optimize significant errors more quickly.

We also utilized the Structural Similarity Index Measure (SSIM) loss [10] to measure the similarity between the predicted depth map $\widehat{D}$ and GT $D$, which focused on preserving structural information. The SSIM loss is particularly effective in capturing perceptual differences, as it considers luminance, contrast, and structural variations, making it useful for depth estimation stages.

$$SSIM(\widehat{D}, D) = \frac{(2u_{\widehat{D}}u_D + C_1)(2\sigma_{\widehat{D}D} + C_2)}{(u_{\widehat{D}}^2 + u_D^2 + C_1)(\sigma_{\widehat{D}}^2 + \sigma_D^2 + C_2)}, \quad (21)$$

where $u_{\widehat{D}}$ and $u_D$ are the means of $\widehat{D}$ and $D$, $\sigma_{\widehat{D}}^2$ and $\sigma_D^2$ are the variances, $\sigma_{\widehat{D}D}$ is the covariance and $C_1$, $C_2$ are the small constants to avoid division by zero.

$$L_{SSIM} = 1 - SSIM(\widehat{D}, D) \quad (22)$$

When the $SSIM$ value was closer to 1, it indicated a high degree of similarity between the predicted depth map $\widehat{D}$ and GT $D$. To convert this similarity measure into a loss for optimization, we computed the 1-SSIM loss. This ensured that minimizing the loss corresponds to maximizing the structural similarity between $\widehat{D}$ and $D$.

In addition, we also adopted a gradient-based loss function to evaluate the boundary consistency and local structural similarity between the predicted depth map $\widehat{D}$ and the GT $D$. Traditional pixel-level losses, such as L1 or L2 losses, primarily focus on the absolute error of pixel values and are insufficient for capturing the boundaries and detailed features in depth maps. To address this limitation, gradient-based loss

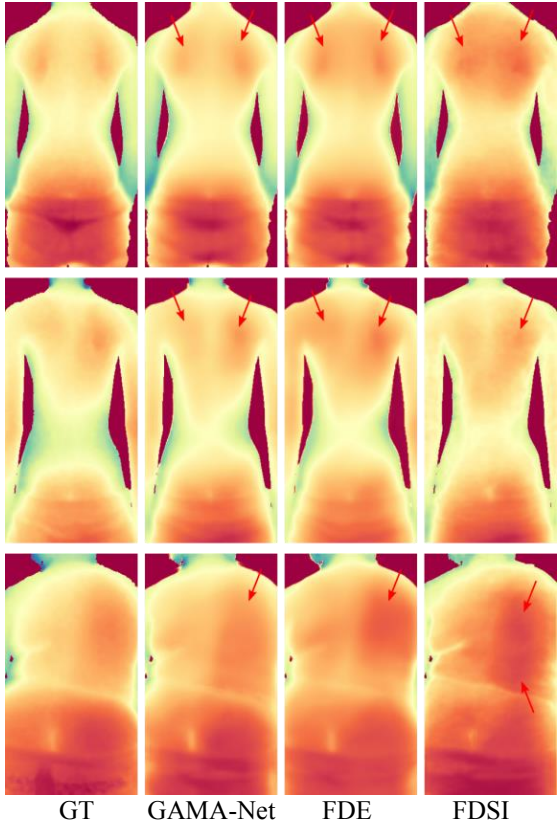

GT     GAMA-Net     FDE     FDSI

**Figure 3.** The visualization comparison of human back depth estimation. Rows (from left to right) represent the GT, the results of our model, FDE, and FDSI, respectively. The arrows highlight areas in the scapular region where GAMA-Net provides more accurate depth estimation.

$$I_{RGBD} = concat(I_{RGB} + D) \quad (18)$$

To be specific, the goal was to predict the curve area of the spine and classify the area at the pixel level. We used a semantic segmentation network U-Net as the main framework for curve generation. The input was an RGBD image $I_{RGBD}$, and the output was the predicted curve segmentation map $M_{curve}$, which was binary segmentation map, indicating the location of the spinal curve.

*4. Loss Function*

The loss function consisted of two components, which supervised the model training for depth estimation and spine curve generation respectively.

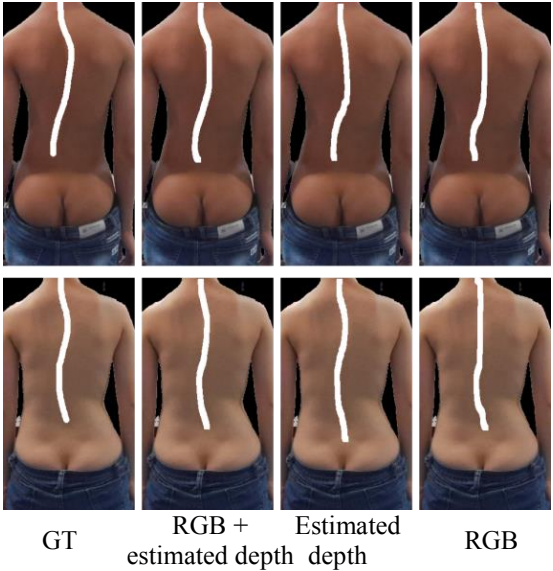

| GT | RGB + estimated depth | Estimated depth | RGB |

**Figure 4.** The visual comparison of spine curve generation using different input data. Rows (from left to right) show the GT spine curve, the generated spine curve using both RGB and estimated depth, using only estimated depth, and using only RGB respectively.

function incorporated depth edge information into the optimization process by calculating the gradients of the depth map, including both horizontal and vertical gradients. It ensured that the model has paid attention to depth variations and boundary details.

$$L_{grad} = \frac{1}{N}\sum_{i=1}^{N}(|\nabla_x \hat{D}_i - \nabla_x D_i| + |\nabla_y \hat{D}_i - \nabla_y D_i|), \quad (23)$$

where $\nabla_x$ and $\nabla_y$ are the gradient of X-axis and Y-axis.

Finally, the total loss was weighted sum of three types of loss, $L_{Berhu}$, $L_{SSIM}$ and $L_{grad}$. We set $\alpha$ as 5, $\beta$ as 1 and $\varepsilon$ as 0.5 respectively.

$$L_{depth} = \alpha L_{Berhu} + \beta L_{SSIM} + \varepsilon L_{grad} \quad (24)$$

*B. Spine Morphology Estimation (Curve Generation) Loss*

We chose cross entropy loss as the loss function of the curve generation stage. The reason for this choice was that we regarded the curve generation problem as a semantic segmentation task, where the goal was to classify each pixel in the image, marking the pixels belonging to the curve area as 1 and the pixels in the background area as 0. Therefore, we chose to use cross entropy loss to optimize, which is described as,

$$L_{curve} = -\frac{1}{N}\sum_{i=1}^{N}\sum_{c=1}^{M} P_{ic} \log(\hat{p}_{ic}), \quad (25)$$

where $N$ is the total number of pixels, $M$ is the total number of classes ($M$=2 for spine curve generation). $p_{ic}$ is the true label of pixel and $\hat{p}_{ic}$ is the predicted probability that pixel $i$ belongs to class $c$.

## III. RESULTS AND DISCUSSION

*A. Depth Estimation*

We validated the effectiveness of the PBHA and AMFF modules in GAMA-Net, as presented in Table 1. Each module

**Table 3.** The quantitative comparison with state-of-the-art methods FDSI and FDE on human back depth estimation.

| Method | Error↓ | | | Accuracy↑ | | |
|---|---|---|---|---|---|---|
| | Abs Rel | RMSE | Sq Rel | $\beta<1.25$ | $\beta<1.25^2$ | $\beta<1.25^3$ |
| FDSI | 0.2143 | 0.1110 | 0.0275 | 0.7324 | 0.9113 | 0.9521 |
| FDE | 0.1956 | 0.0662 | 0.0149 | 0.7661 | 0.9253 | 0.9673 |
| **Our** | **0.1887** | **0.0638** | **0.0140** | **0.7815** | **0.9356** | **0.9747** |

*Abs Rel: absolute relative error, RMSE: root mean square error, Sq Rel: squared relative error.

demonstrates significant contributions to the model's performance. GAMA-Net achieves a nearly 4% improvement in accuracy under strictest measurement standard compared to the baseline. Additionally, using the AMFF module alone results in approximately a 1% improvement, while incorporating the PBHA module alone leads to a roughly 2% enhancement. These improvements highlight the impact of the PBHA and AMFF modules, which are specifically designed to capture finer details and enhance the precision of depth estimation.

To evaluate the contributions of individual components in our composite loss function, we conducted a series of ablation experiments, as detailed in Table 2. The results demonstrate that each loss component contributes to optimizing the model's performance. Combining all components into a composite loss function yields the best overall performance, as it effectively emphasizes finer details, similarity, and boundary consistency.

We also compared our model GAMA-Net with the-state-of-the-art methods, FDSI [23] and FDE [25]. Table 3 presents a quantitative comparison of their performance in estimating relative depth values (normalized to the range [0, 1]). GAMA-Net shows superior performance by effectively capturing fine details. In addition to quantitative evaluation, we also validated the model's visual performance, as shown in Figure 3. The visualizations highlight GAMA-Net's ability to accurately capture subtle depth differences in localized regions of the back, enabling finer depth reconstruction. These fine-grained depth variations are critical for subsequent spine curve generation.

*B. Spine Morphology Estimation (Curve generation)*

To validate the effectiveness of depth map for spine morphology estimation, we used it to generate spine curves. Specifically, we combined the estimated depth map with the RGB image to create a 4-channel RGBD image. As shown in Table 4, using RGBD images as input enables the model to produce more accurate spine curves compared to using only depth or RGB images. Such discrepancy arises because the depth map lacks rich texture, color, and other two-dimensional visual cues, while the RGB image lacks geometric information. These limitations reduce the model's accuracy in segmenting the spinal region, ultimately impacting the precision of curve generation. Figure 4 further illustrates the predicted curves generated using RGBD, RGB, and depth map inputs, along with the GT curve. The results highlight the superiority of RGBD images in analyzing spine morphology, underscoring the importance of incorporating depth estimation in related research.

**Table 4.** The quantitative comparison on curve generation performance using estimated depth, RGB, and RGB with estimated depth inputs.

| Method | IOU | Accuracy | Dice Similarity | Sensitivity | Specificity |
|---|---|---|---|---|---|
| Estimated depth | 0.7914 | 0.9652 | 0.8820 | 0.9530 | 0.9672 |
| RGB | 0.8025 | 0.9678 | 0.8887 | 0.9421 | 0.9719 |
| **RGBD** | **0.8102** | **0.9695** | **0.8939** | **0.9411** | **0.9740** |

IV. CONCLUSION

In conclusion, this paper proposes GAMA-Net, a novel framework for accurate human back depth information from single RGB images, aimed at enhancing spine morphology analysis. The proposed dual-feature extractor architecture, incorporating both global and patch-based encoders, was coupled with specialized attention (PBHA) and feature-fusion (AMFF) modules that effectively capture fine-grained details while maintaining a global perspective. Experimental results show that GAMA-Net delivers superior depth estimation, particularly in regions with subtle depth variations. Integrating the predicted depth with RGB images to form RGBD inputs, significantly improves 3D spine curve generation, highlighting the value of geometric information in spine assessments. This research represents one of the first high-precision depth estimation methods for human back, offering a radiation-free and accessible solution for scoliosis screening and spine health evaluations.

ACKNOWLEDGMENT

This work was supported by Department Seed Fund of The University of Hong Kong (200011026), Seed Fund for Basic Research for New Staff of The University of Hong Kong (103034011), The National Natural Science Foundation of China Young Scientists Fund (NSFC 82402398).

REFERENCES

[1] P. Dou, X. Li, H. Jin, B. Ma, M. Jin, and Y. Xu, "Research trends of biomechanics in scoliosis from 1999 to 2023: a bibliometric analysis," Spine deformity, 2024.
[2] M. Li, Q. Nie, J. Liu, and Z. Jiang, "Prevalence of scoliosis in children and adolescents: a systematic review and meta-analysis,", Frontiers in Pediatrics, vol. 12, p. 1399049, 2024.
[3] M. Zhao, N. Meng, J. P. Y. Cheung, C. Yu, P. Lu, and T. Zhang, "SpineHRformer: A Transformer-Based Deep Learning Model for Automatic Spine Deformity Assessment with Prospective Validation," Bioengineering, vol. 10, no. 11, p. 1333, 2023.
[4] N. Kontodimopoulos, K. Damianou, E. Stamatopoulou, A. Kalampokis, and I. Loukos, "Children's and parents' perspectives of health-related quality of life in newly diagnosed adolescent idiopathic scoliosis," Journal of Orthopaedics, vol. 15, no. 2, pp. 319-323, Jun. 2018.
[5] N. Meng et al., "An artificial intelligence powered platform for auto-analyses of spine alignment irrespective of image quality with prospective validation," eClinicalMedicine, vol. 43, 2022.
[6] N. Meng et al., "EUFormer: Learning Driven 3D Spine Deformity Assessment with Orthogonal Optical Images," in 2024 46th Annual International Conference of the IEEE Engineering in Medicine and Biology Society (EMBC), 2024: IEEE, pp. 1-4.
[7] T. Zhang et al., A clinical classification for radiation-less monitoring of scoliosis based on deep learning of back photographs. 2022.
[8] S. M. Presciutti, T. Karukanda, and M. Lee, "Management decisions for adolescent idiopathic scoliosis significantly affect patient radiation exposure," The Spine Journal, vol. 14, no. 9, pp. 1984-1990, Sep. 2014.
[9] C. L. Nash, Jr., E. C. Gregg, R. H. Brown, and K. Pillai, "Risks of exposure to X-rays in patients undergoing long-term treatment for scoliosis," J Bone Joint Surg Am, vol. 61, no. 3, pp. 371-4, Apr. 1979.
[10] N. Meng, H. K.-H. So, X. Sun, and E. Y. Lam, "High-dimensional dense residual convolutional neural network for light field reconstruction," IEEE transactions on pattern analysis and machine intelligence, vol. 43, no. 3, pp. 873-886, 2019.
[11] N. Meng, K. Li, J. Liu, and E. Y. Lam, "Light field view synthesis via aperture disparity and warping confidence map," IEEE Transactions on Image Processing, vol. 30, pp. 3908-3921, 2021.
[12] N. Meng, X. Wu, J. Liu, and E. Lam, "High-order residual network for light field super-resolution," in Proceedings of the AAAI Conference on Artificial Intelligence, 2020, vol. 34, no. 07, pp. 11757-11764.
[13] T. Zhang et al., "Deep Learning Model to Classify and Monitor Idiopathic Scoliosis in Adolescents Using a Single Smartphone Photograph," JAMA Network Open, vol. 6, no. 8, pp. e2330617-e2330617, 2023.
[14] X. Zhu et al., "MGScoliosis: Multi-grained scoliosis detection with joint ordinal regression from natural image," Alexandria Engineering Journal, vol. 111, pp. 329-340, Jan. 2025.
[15] T. Kokabu et al., "An algorithm for using deep learning convolutional neural networks with three-dimensional depth sensor imaging in scoliosis detection," The Spine Journal, vol. 21, no. 6, pp. 980-987, Jun. 2021.
[16] Y. Liang et al., "3D Spine Model Reconstruction Based on RGBD Images of Unclothed Back Surface," IEEE Transactions on Biomedical Engineering, vol. 71, no. 1, pp. 270-281, 2024.
[17] C. Yang, M. Chen, H. Xu, J. Li, and Q. Huang, "Fully automatic spinal scanning and measurement based on multi-source vision information," Measurement, vol. 224, p. 113955, Jan. 2024.
[18] Y. Ming, X. Meng, C. Fan, and H. Yu, "Deep learning for monocular depth estimation: A review," Neurocomputing, vol. 438, pp. 14-33, May. 2021.
[19] A. Mertan, D. J. Duff, and G. Unal, "Single image depth estimation: An overview," Digital Signal Processing, vol. 123, p. 103441, Apr. 2022.
[20] N. Chong and F. Yang, "A monocular medical endoscopic images depth estimation method based on a confidence-guided dual-branch siamese network," Biomedical Signal Processing and Control, vol. 102, p. 107123, Apr. 2025.
[21] B. H. Jeong, H. K. Kim, and Y. D. Son, "Depth estimation from monocular endoscopy using simulation and image transfer approach," Computers in Biology and Medicine, vol. 181, p. 109038, Oct. 2024.
[22] B. Huang et al., "Simultaneous Depth Estimation and Surgical Tool Segmentation in Laparoscopic Images," IEEE Transactions on Medical Robotics and Bionics, vol. 4, no. 2, pp. 335-338, 2022.
[23] F. Khan, W. Shariff, M. A. Farooq, S. Basak, and P. Corcoran, "A Robust Light-Weight Fused-Feature Encoder-Decoder Model for Monocular Facial Depth Estimation From Single Images Trained on Synthetic Data," IEEE Access, vol. 11, pp. 41480-41491, 2023.
[24] F. Khan, S. Basak, H. Javidnia, M. Schukat, and P. Corcoran, "High-accuracy facial depth models derived from 3D synthetic data," in 2020 31st Irish Signals and Systems Conference, 2020: IEEE, pp. 1-5.
[25] F. Khan, S. Hussain, S. Basak, J. Lemley, and P. Corcoran, "An efficient encoder–decoder model for portrait depth estimation from single images trained on pixel-accurate synthetic data," Neural Networks, vol. 142, pp. 479-491, Oct. 2021.
[26] C. Zhao, Q. Sun, C. Zhang, Y. Tang, and F. Qian, "Monocular depth estimation based on deep learning: An overview," Science China Technological Sciences, vol. 63, no. 9, pp. 1612-1627, 2020.
[27] J. Wang et al., "Deep High-Resolution Representation Learning for Visual Recognition," IEEE Transactions on Pattern Analysis and Machine Intelligence, vol. 43, no. 10, pp. 3349-3364, 2021.
[28] X. Jia, Q. Lin, and W. Ding, "An ultra-high-definition multi-exposure image fusion method based on multi-scale feature extraction," Applied Soft Computing, vol. 166, p. 112240, Nov. 2024.
[29] H. Xing, W. Wei, L. Zhang, and Y. Zhang, "Multi-scale feature extraction and fusion with attention interaction for RGB-T tracking," Pattern Recognition, vol. 157, p. 110917, Jan. 2025.
[30] N. Mungoli, "Adaptive feature fusion: enhancing generalization in deep learning models," arXiv preprint arXiv:2304.03290, 2023.
[31] L. Zwald and S. Lambert-Lacroix, "The berhu penalty and the grouped effect," arXiv preprint arXiv:1207.6868, 2012.